# JClarens: A Java Framework for Developing and Deploying Web Services for Grid Computing

Michael Thomas,[1] Conrad Steenberg,[1] Frank van Lingen,[1] Harvey Newman,[1]
Julian Bunn,[1] Arshad Ali,[2] Richard McClatchey,[3] Ashiq Anjum,[2,3] Tahir Azim,[2]
Waqas ur Rehman,[2] Faisal Khan,[2] Jang Uk In[4]

[1]California Institute of Technology, Pasadena, CA, USA
{thomas,conrad,newman}@hep.caltech.edu, {fvlingen, Julian.Bunn}@caltech.edu
[2]National University of Sciences and Technology, Rawalpindi, Pakistan
{arshad.ali,ashiq.anjum,tahir.azim,waqas.rehman,faisal.khan}@niit.edu.pk
[3]University of the West of England,Bristol ,UK  Richard.McClatchey@uwe.ac.uk
[4]University of Florida, Gainsville, FL, USA   juin@phys.ufl.edu

## Abstract

*High Energy Physics (HEP) and other scientific communities have adopted Service Oriented Architectures (SOA) [1][2] as part of a larger Grid computing effort. This effort involves the integration of many legacy applications and programming libraries into a SOA framework. The Grid Analysis Environment (GAE) [3] is such a service oriented architecture based on the Clarens Grid Services Framework [4][5] and is being developed as part of the Compact Muon Solenoid (CMS) [6] experiment at the Large Hadron Collider (LHC) [7] at European Laboratory for Particle Physics (CERN) [8]. Clarens provides a set of authorization, access control, and discovery services, as well as XMLRPC and SOAP access to all deployed services. Two implementations of the Clarens Web Services Framework (Python and Java) offer integration possibilities for a wide range of programming languages. This paper describes the Java implementation of the Clarens Web Services Framework called 'JClarens.' and several web services of interest to the scientific and Grid community that have been deployed using JClarens.*

## 1. Introduction

These days scientific collaborations require more computing (cpu, storage, networking etc) resources than can be provided by any single institution. Furthermore, these collaborations consist of many geographically dispersed groups of researchers. As an example, two High Energy Physics experiments CMS [6] and ATLAS [9] will be generating petabytes to exabytes of data that must be accessible to over 2000 physicists from 150 participating institutions in more than 30 countries.

Grid computing holds the promise of harnessing computing resources at geographically dispersed institutions into a larger distributed system that can be utilized by the entire collaboration.

As one of any site's responsibilities is to the local users, institutes managing these sites generally have a large amount of control over their site's resources. Local requirements at each site lead to different decisions on operating systems, software toolkits, and usage policies. These differing requirements can result in a very heterogeneous character of some Grid environments.

A Services Oriented Architecture (SOA) is well suited to addressing some of the issues that arise from such a heterogeneous, locally controlled but globally shared system. Three features of a SOA make it a suitable candidate for use in a Grid computing environment:

**Standard Interface Definitions** – The use of common interface definitions for resources allows them to be used in the same way.

**Implementation Independence** – Any programming language on any operating system can be used to implement services. Local sites are not required to run a particular operating system or to use a specific programming language when implementing or selecting their service implementations.

**Standard Foundation Services** – These services provide the basic security, discovery, and access control features for locating and accessing remote services.

One example of a SOA is the Grid Analysis Environment (GAE) [3] which uses the Clarens Grid Service Framework. The GAE aims to provide a coherent environment for thousands of physicists to utilize the geographically dispersed computing resources for data analysis. The GAE SOA is part of the Ultralight [10] project which focuses on integration of networks as an end-to-end, managed resource in global, data-intensive Grid systems. Clarens is also utilized as a Grid Service Framework in other projects, such as Lambda station [11], Proof Enabled Analysis Center (PEAC) [12], and Physh [13].

In this paper, we discuss the design, implementation and various services of the second Clarens implementation called 'JClarens.' The JClarens architecture is based on Java Servlet Technology and XMLRPC/SOAP. JClarens can be deployed with any web server configured with a servlet engine. Both JClarens and PClarens (based on Python) are intended to be deployed as a set of peer-to-peer servers with a complementary set of deployed services.

## 2. Software architecture

The core of the Clarens Framework provides a set of standard foundation services for authorization, access control, service discovery and a framework for hosting additional grid services. These additional services offer functionality, such as remote job submission, job scheduling, data discovery and access. Standard technologies such as Public Key Infrastructure (PKI) for security, and SOAP/XMLRPC for invoking remote services, are used within the Clarens framework to provide secure and ubiquitous access. The two implementations of Clarens (Python and Java) both share the common set of standard foundation services. PClarens, is based on the Apache Web Server with the mod_python module. In this implementation services are written using the Python programming language.

JClarens is developed as a single J2EE web application hosted inside the Tomcat servlet container [14]. Tomcat provides the basic HTTP transport for sending and receiving the SOAP and XMLRPC messages. The HTTPS transport is also supported. The built-in HTTPS connector for Tomcat could not be used, however, due to its lack of support for client authentication using proxy certificates. Libraries provided by the gLite [15] project provide support for this client proxy authentication. HTTP-based authentication can also be performed using the 'system' service described below.

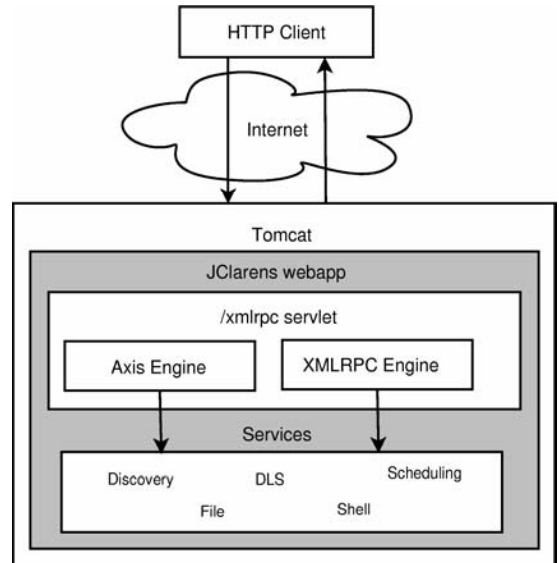

**Figure 1. JClarens architecture**

The JClarens web application uses a single servlet to handle both SOAP and XMLRPC messages, as shown in Figure 1. The servlet contains an instance of both an Axis SOAP engine and an Apache XML-RPC engine. Each incoming request is examined to determine if it is a SOAP or XMLRPC request. Requests are then passed to the appropriate engine (Axis or XML-RPC) for processing. The use of a single URL for both types of requests simplifies the configuration for client applications.

The use of two transport encoding engines necessitates two configurations. A Java properties file is used to configure the main JClarens web application, as well as the XMLRPC engine. The interface for a web service is specified using one configuration property, and another property is used to define the implementation class for that service. This makes it very easy to install multiple service implementations, and choose between the multiple implementations at startup time. A standard SOAP web service deployment descriptor file is used to configure the SOAP engine.

# 3. Services

In a dynamic Grid environment, there is no central authority that can be trusted for all authorization and access control queries. Each local provider of a Grid service is responsible for providing these features for their set of local services. The Clarens framework provides several core services [16], which includes a system service that provides authorization and access control management capabilities. A group service augments the access control capabilities by allowing groups of users to be managed as a single entity. The file service provides a limited mechanism for browsing, uploading, and downloading files on a remote system. A proxy service assists in the management of client proxy credentials that are used when executing shell commands (*see* Shell Service, Part 3.4, *infra*) and when making delegated service calls from one service to another.

The core services described in the previous paragraph are part of the standard installation of Clarens. As Clarens is continuously being developed and used within projects, new functionalities and services are being created. Rather than creating a large package containing all the possible services, site administrators can decide to install additional services

Additional services add useful functionality for operating in a Grid environment as well as some domain-specific services for use in very particular environments such as CMS. The next sections discuss several of these additional services.

## 3.1. Job scheduling

The job of a Grid scheduler is to enable effective use of resources at many Grid sites. The scheduler takes into account data location, CPU availability, and resource policies to perform a matchmaking process between the user's request and the Grid resources. SPHINX [17][18] is a novel scheduling middleware in a dynamically changing and heterogeneous Grid environment and provides several key functionalities in its architecture for efficient and fault tolerant scheduling in dynamic Grid environments.

*Modular system:* The scheduling system can be easily modified or extended. Changes to an individual scheduling module can be made without affecting the logical structure of other modules.

*Robust and recoverable system:* The SPHINX server uses a relational database to manage the scheduling process. The database also provides a level of fault tolerance by making the system easily recoverable from internal component failures.

*Platform independent interoperable system:* The JClarens service framework is used by SPHINX to provide platform and language neutral XML-based communication protocols.

SPHINX consists of two components: a client and a server. This separation facilitates system accessibility and portability. The client is a lightweight, portable scheduling agent that provides the server with scheduling requests. The client also interacts with a user which submits scheduling requests to the client. As such, it provides an abstract layer to the scheduling service, while exposing a customized interface to accommodate user specific functions.

SPHINX processes Directed Acyclic Graphs (DAGs) using a system based on a finite state machine. Each DAG workflow (and corresponding set of jobs) can be in one of several states. These states allow for an efficient control flow as well as graceful recovery in the case of machine failure. Each state has a module to handle control flow and machine failure. The data management component communicates with the monitoring interface and a Replica Location Service (RLS) for managing copies of data stored at many sites.

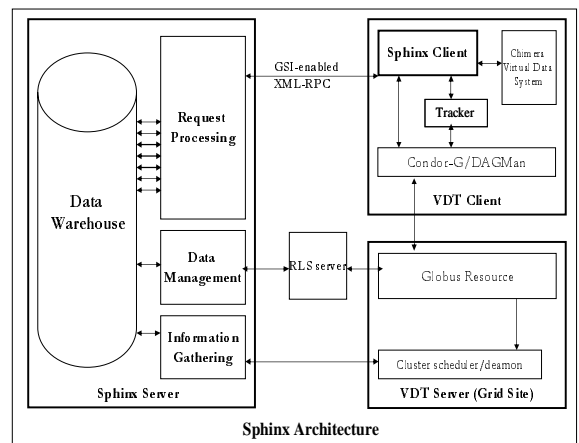

**Figure 2. SPHINX architecture**

Figure 2 shows the architecture of SPHINX. The SPHINX client forwards the job generated by Chimera [19] to the server for execution site recommendation. The SPHINX server schedules the job on to a site utilizing the monitoring information and Replica Location Service. The client can then submit it to that site and the tracker can send back job status information to the server. The communication between all the components uses GSI-enabled XML-RPC services.

## 3.2. Discovery Service

Within a global distributed service environment services will appear, disappear, and be moved in an unpredictable and dynamic manner. It is virtually impossible for scientists and applications to keep track of these changes. The *discovery service* provides scientists and applications the ability to query for services and to retrieve up-to-date information on the location and interface of a service in a dynamic environment. Although the discovery service is conceptually similar to a UDDI registry, it offers a much simpler service interface and more dynamic content. Registration with the *discovery service* must happen at regular intervals in order to prove that a service is still available. If a service fails to notify the *discovery service* within a certain time period, it is automatically removed from the registry. The *discovery service* offers four methods: (1) `register` is used to add a new service to the registry; (2) `find_server` is used to locate service hosts that match certain search criteria; (3) `find` is used to locate service instances that match certain search criteria; (4) `deregister` is used to remove services from the registry (however it is seldom used since the registry will automatically remove the service once it fails to re-register).

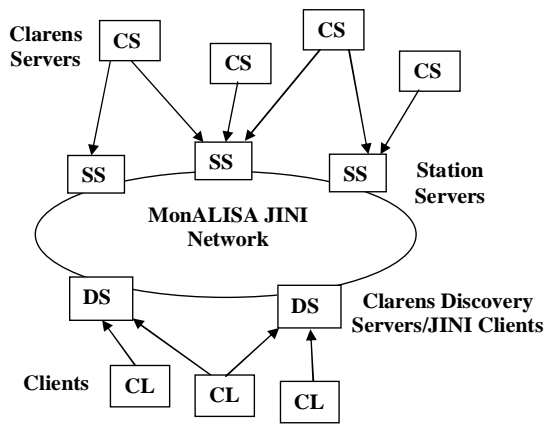

**Figure 3. Discovery Service**

Two implementations of the *discovery service* interface are currently in use. The first is a peer-to-peer implementation based on Jini and the MonALISA [20] Grid monitoring system, as shown in Figure 3. MonALISA is a Jini-based monitoring system that uses station servers to collect local monitoring data, and uses a Jini peer-to-peer network to share selected monitoring data with other station servers or other interested clients. Arbitrary monitoring data can be published to a MonALISA station server using ApMon, a library that uses simple XDR encoded UDP packets.

This first implementation of the *discovery service* uses the ApMon library to publish service registrations to the MonALISA Jini network. Each *discovery service* contains a client that listens for these service publications on the MonALISA Jini network, and stores them in an in-memory cache. The *discovery service* periodically purges expired entries from this in-memory cache. Since the service registry is stored in memory, it is not persistent across server restarts, which is not a problem since the registry will be quickly populated with new information once it starts up again.

The second implementation of the *discovery service* does not make use of the MonALISA Jini network at all. Instead, it uses a UDDI repository to store the service publications. The UDDI implementation serves as an important bridge between the simple JClarens service registry and a more full-featured and common UDDI registry. The `register` method of this implementation makes the appropriate UDDI service calls to insert the Web Service into a known UDDI registry. The `find` and `find_server` methods perform queries on the UDDI registry and reformat the result to match the *discovery service* interface. This UDDI implementation allows JClarens to be used in an environment that is more static and has existing UDDI repositories already in use.

## 3.3. Data Location Service

The Grid will have a number of computational elements that might belong to multiple organizations split across geographically dispersed sites. The sites are connected by a number of different Wide Area Network (WAN) links. These links will have different bandwidths and latencies for various reasons such as the relative locations of the sites, the capabilities of the local telecommunications providers, etc. Some executables and data items will be large compared to the available network bandwidths and latencies. The relative locations of executables and data, network bandwidth consumed, size and time of the data transfer are important within Grid wide scheduling decisions as these parameters might have significant impact on the computing costs.

In order to achieve some of the objectives stated above, File Access and optimized Replica Location Services in a Grid analysis environment must be combined so that a user or user agent can specify a single logical filename and return the optimal physical path to that file. The Data Location Service (DLS) outlined in this paper focuses on the selection of the best replica of a selected logical file, taking into account the location of the computing resources and network and storage access latencies. It must be a light-weight web service that gathers information from the Grid's network monitoring service and performs access optimization calculations based on this information.

This service provides optimal replica information on the basis of both faster access and better performance characteristics. The Data Location process allows an application to choose a replica, from among those in various replica catalogs, based on its performance and data access features. Once a logical file is requested by the user, the DLS uses the replica catalog to locate all replica locations containing physical file instances of this logical file, from which it should choose an optimal instance for retrieval. These decisions are based on criteria such as network speed, data location, file size, data transfer time and other related parameters. It should be decentralized (not rely on some central storage system) and fault tolerant, so that when one instance goes offline, the user (or client service) is still able to work by using other instances of the service.

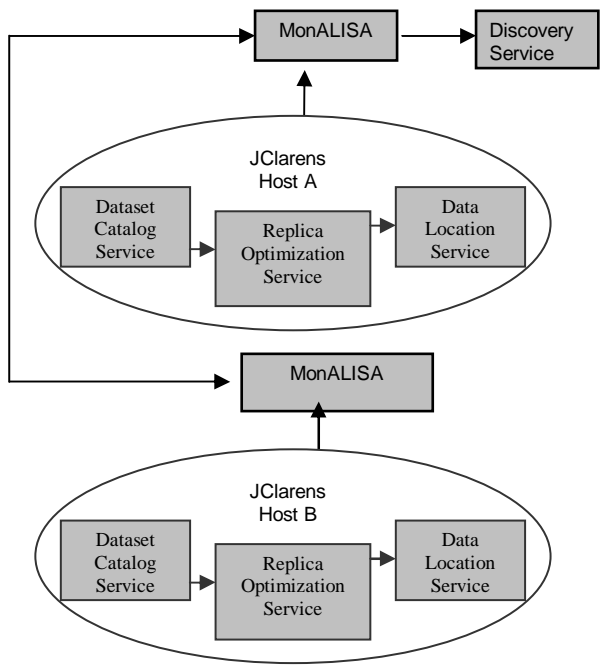

**Figure 4. DLS Service Architecture**

Figure 4 shows the DLS architecture. The DLS is a decentralized service which takes into account the selection process on the basis of client and file location and respective network parameters while utilizing the *discovery service* to locate the available *dataset catalog* services. Each catalog is queried by the DLS to find all locations where the requested file is available. The service returns a paginated list of file locations to the caller. In addition, the DLS monitors data access patterns, keeping track of often-used and often-unavailable files.

The result of a call to this service is sorted by either the reliability of the file, or by the "closeness" determined by some network ping time or other network measurements from MonALISA. The DLS also evaluates the network costs for accessing a replica. For this it must use information such as estimates of file transfer times based on network monitoring statistics. The DLS selects the "best" physical file based on the given parameters. Several performance parameters such as network speed, current network throughput, load on data servers, input queues on data server are included in the metric for selecting the "best" replica. The *discovery service* makes it possible to discover and publish services in a fault tolerant and decentralized way.

### 3.4. Shell Service

The *shell service* provides a generic way to invoke local shell commands using web service calls in a secure manner. A Globus grid-mapfile is used to map a user's x509 credentials to a local user account. The *shell service* makes use of the `suexec` command line tool from Apache in order to change to the local user's uid before executing the command. `suexec` has built-in safeguards to ensure that it can not be used to run commands as root or any other privileged system user.

The core method in the *shell service* is the `cmd` method. This is the method that is used to execute the shell commands on the JClarens server. It assigns a unique ID to each request and returns this ID back to client after the requested command has been scheduled on the server. This command allows the client to launch long-running commands without having to maintain a persistent connection to the server. The *shell service* also provides the `cmd_info` method for obtaining information on the status of the scheduled commands. This information includes the name of the local system user that was used to run the command, the process ID, start and end times of the command's execution, and directory containing the sandbox used as the working directory for the command. This directory stores standard output and error files that contain any output and errors from the command. The previously mentioned *file service* can be used to retrieve contents of these files.

The ability to execute arbitrary commands as a local system user makes the *shell service* a valuable tool for integrating existing command line tools into web services. By providing a set of utility functions in the JClarens core server, new web services can be written that simply wrap command line tools and return the output to the user. For example, a *df service* can be written using the *shell service* utilities that executes the system `df` command and returns the disk usage statistics back to the user.

## 4. Adding new services

One of the primary motivations for creating JClarens is to assist web service developers with the development and deployment of new services while providing functionality needed by all services deployed in a dynamic Grid environment: authorization, authentication, and discoverability. JClarens uses the common Redhat Package Manager (RPM) packaging format for installation. This RPM package comes with a Tomcat 5.0.28 servlet engine as well as the JClarens web application itself. Once the RPM package has been installed, no further configuration is necessary to start the server. Some site-specific configuration (such as the use of a host-specific x509 certificate) will be necessary before putting the server into production.

```
# rpm -ivh jclarens-0.5.2-2.i386.rpm
```
**Example 1. Command to install JClarens**

A sample service is provided as a template for building new services. This sample service contains an Ant[21]-based build file with targets to generate SOAP stubs from WSDL. The build file also contains rules for building an RPM for the service. This RPM can then be installed into an existing JClarens installation. After installing the service RPM, only a single line need be added to the global JClarens configuration file to indicate that this service should now be used. The sample service also contains a template XMLRPC binding file that can be easily customized to enable XMLRPC encoding for the service.

```
# rpm -ivh jclarens-shell-0.5.2-2.i386.rpm
```
**Example 2. Command to install a new service**.

The process to deploy a new service in JClarens is slightly more involved than with a basic Apache Axis SOAP engine. However, JClarens provides more features than the basic Axis engine, These features include the dynamic service publication and XMLRPC bindings.

## 5. Future Developments

The JClarens Web Services Framework is becoming a foundation for web service development and deployment in several projects such as GAE that focus on distributed and scalable scientific analysis within the CMS experiment. Future work will focus on developing services to create a self-organizing, autonomous Grid.

One such service currently under development is the *job monitoring service* [22] that will be tightly coupled with a Grid scheduler. This service will provide the user with a view of the current status of their job submission request. A set of estimation service methods [23] will determine how long a particular Grid job submission (be it a large data transfer request or some long-running computation) will take. A *steering service* [24] will provide means for the user to fine-tune a job submission, so that he can redirect slow-running jobs to faster computing sites. Future improvements on the *steering service* will add more autonomous behavior, making use of the *job monitoring service* and estimation methods to automatically detect when job execution could be optimized (reducing execution time or resource usage) and steering such jobs automatically on behalf of the user.

Additionally, improvements will be made to further simplify the process of writing and deploying new services. A tool that will automatically generate XMLRPC binding classes from the service WSDL description is being investigated. This would reduce the amount of code that a service author needs to write so that she doesn't need to even know that the SOAP or XMLRPC bindings exist.

Most services require some sort of database connectivity. JClarens supports many databases through the use of JDBC, but does so in a somewhat inelegant manner. Currently each supported database requires a new implementation of the service interface due to non-portable table-creation statements. A database abstraction layer would help remove some of these database-specific features, minimizing the need for multiple service implementations.

## 6. Related work

There are many other international Grid projects underway in other scientific communities. These can be categorized as integrated Grid systems, core and user-level middleware, and application-driven efforts. Some of these are customized for the special requirements of the HEP community. Others do not accommodate the data intensive nature of the HEP Grids and focus upon the computational aspect of Grid computing.

EGEE [2] middleware, called gLite [15], is a service-oriented architecture. The gLite Grid services aim to facilitate interoperability among Grid services and frameworks like JClarens and allow compliance with standards, such as OGSA [25], which are also based on the SOA principles.

Globus [26] provides a software infrastructure that enables applications to handle distributed heterogeneous computing resources as a single virtual machine. Globus provides basic services and capabilities that are required to construct a computational Grid. Globus is constructed as a layered architecture upon which the higher-level JClarens Grid services can be built.

Legion [28] is an object-based "meta-system" that provides the software infrastructure so that a system of heterogeneous, geographically distributed, high-performance machines can interact seamlessly. Several of the aims and goals of both projects are similar but compared to JClarens the set of methods of an object in Legion are described using Interface Definition Language.

The Gridbus [28] toolkit project is engaged in the design and development of cluster and Grid middleware technologies for service-oriented computing. It uses Globus libraries and is aimed for data intensive sciences and these features make Gridbus conceptually equivalent to JClarens.

NASA's IPG [29], is a network of high performance computers, data storage devices, scientific instruments, and advanced user interfaces. Due to its Data centric nature and OGSA compliance, IPG services can potentially interoperate with GAE services.

WebFlow [30] framework for the wide-area distributed computing. is based on a mesh of Java-enhanced Apache web servers, running servlets that manage and coordinate distributed computation and it is architecturally closer to JClarens .

NetSolve [31] system is based around loosely coupled, distributed systems, connected via a LAN or WAN. Netsolve clients can be written in multiple languages as in JClarens and server can use any scientific package to provide its computational software.

The Gateway system offers a programming paradigm implemented over a virtual web of accessible resources [32].Although it provides a portal behaviour like JClarens and is based on SOA, its design is not intended to support data intensive applications.

The GridLab [33] will produce a set of Grid services and toolkits providing capabilities such as dynamic resource brokering, monitoring, data management, security, information, adaptive services and more. GAE Services can access and interoperate with GridLab services due to its SOA based nature.

The Open Grid Services Architecture (OGSA) framework, the Globus-IBM vision for the convergence of web services and Grid computing has been taken over by Web Services Resource Framework (WSRF) [34]. WSRF is inspired by the work of the Global Grid Forum's Open Grid Services Infrastructure (OGSI) [35]. The developers of the Clarens Web Services Framework are closely following these developments.

## 7. Conclusion

JClarens has proven to be an important component of the Grid Analysis Environment. As a second implementation of the Clarens Grid Services Framework, it has provided a way to integrate existing Java Grid services with little difficulty. The tight integration with the MonALISA monitoring system has given rise to a new set of Grid services that can use a global view of the state of the Grid in order to make optimized decisions.

JClarens was chosen as a Grid service host by the SPHINX development team based on JClarens' Java implementation, its MonALISA integration, and its publication of services via the dynamic *discovery service* service registry. Already more projects within the HEP community are looking to JClarens to host their Grid services.

## 8. Acknowledgements

This work is partly supported by the Department of Energy grants: DE-FC02-01ER254559, DE-FG03-92-ER40701, DE-AC02-76CH03000 as part of the Particle Physics DataGrid project and by the National Science Foundation grants: ANI-0230967, PHY-0303841, PHY-0218937, PHY-0122557. Any